\documentclass[3p,times,procedia]{elsarticle}
\usepackage{nupha_ecrc}

\volume{00}

\firstpage{1}

\journalname{Nuclear Physics A}

\runauth{}

\jid{nupha}

\jnltitlelogo{Nuclear Physics A}

\usepackage{amssymb}

\usepackage[figuresright]{rotating}

\begin{document}

\begin{frontmatter}



\dochead{}

\title{The topological structures in strongly coupled QGP with
chiral fermions on the lattice}

 \author[label1]{Sayantan Sharma}
 \author[label2]{Viktor Dick}
 \author[label1,label2]{Frithjof Karsch}
 \author[label2]{Edwin Laermann}
 \author[label1]{Swagato Mukherjee}
\address[label1]{Physics Department, Brookhaven National Laboratory, \\
        Upton, New York-11973\\}
\address[label2]{Fakult\"{a}t f\"{u}r Physik, Universit\"{a}t Bielefeld
\\        Universit\"{a}tstasse 25, D33619 Bielefeld}

%

\begin{abstract}
The nature of chiral phase transition for two flavor QCD is an interesting but unresolved problem. One of the most intriguing issues is
whether or not the anomalous U(1) symmetry in the flavor sector is effectively restored along with the chiral symmetry.
 This may determine the universality class of the chiral phase transition. Since the physics near the chiral phase transition is essentially
non-perturbative, we employ first principles lattice techniques to address this issue. We use overlap fermions, which have exact chiral
symmetry on the lattice, to probe the anomalous U(1) symmetry violation of  2+1 flavor dynamical QCD configurations with domain wall 
fermions. The latter also optimally preserves chiral and flavor symmetries on the lattice, since it is known that the remnant chiral 
symmetry of the light quarks influences the scaling  of the chiral condensate in the crossover transition region. We observe that the 
anomalous U(1) is not effectively restored in the chiral crossover region.  We perform a systematic study of the finite size and cut-off
effects since the signals of U(1) violation are sensitive to it.
We also provide a glimpse of the microscopic topological structures of the QCD medium that are responsible for the strongly interacting
nature of the quark gluon plasma phase. We study the effect of these microscopic constituents through our first calculations for the
topological susceptibility of QCD at finite temperature, which could be a crucial input for the equation of state for anomalous 
hydrodynamics.

\end{abstract}


\end{frontmatter}


\section{Introduction}
\label{intro}
Understanding the vacuum structure of QCD is one of the most interesting and intriguing 
field of research for about thirty years now. New signals of non-trivial topological fluctuations in QCD 
have been proposed in recent times, like chiral magnetic effect~\cite{chiralm} and anomaly induced 
transport~\cite{chiralt}.  It is an ongoing topic of research in trying to understand appropriate experimental 
signatures of chiral magnetic effect in heavy ion collision experiments at RHIC at BNL and at CERN.
Recent studies of hydrodynamic evolution in presence of chiral anomaly has also provided hints 
towards charge separation of different pion species~\cite{hirono}. However this is not a unique signature of 
non-trivial topological fluctuations in QCD. Effects of topological fluctuations would 
also show up in ratios of thermodynamic observables near the chiral phase transition at vanishingly 
small baryon densities if indeed the strongly interacting matter created in heavy-ion collision thermalizes.
As known from the pioneering paper~\cite{pw} that $U_A(1)$ though an anomalous symmetry in QCD,  
may affect the nature of phase transition in QCD with two light quark flavors. If the $U_A(1)$ anomaly 
is not effectively restored at the chiral crossover transition then in the limit of vanishingly small 
light quark mass, the phase transition is second order with $O(4)$ critical exponents, otherwise it could 
be either first order or second order with $U(2)\times U(2)$ universality class~\cite{bpv}. Remarkably, it is known from model 
quantum field theories in the $O(4)$ universality class that the ratio of higher order baryon fluctuations 
like $\chi_6^B/\chi_2^B$ should have a very characteristic behavior at the chiral crossover transition~\cite{karsch}. 
Measuring such observables in experiments would allow us to understand whether $U_A(1)$ is still effectively 
broken and thus indirectly find out the topological fluctuations responsible for it. Our earlier work~\cite{viktor} has lead to 
the hints that $U_A(1)$ may not be effectively restored at $1.5 ~T_c$, $T_c$ is the chiral crossover transition 
temperature and weakly interacting instantons may be responsible for its breaking at such temperatures. In this talk, 
I present our latest results on our ongoing studies about understanding the mechanism for the $U_A(1)$ breaking at 
temperatures between $1$-$1.2~T_c$. We investigate the nature of the topological structures that cause $U_A(1)$ breaking 
at these temperatures from the eigenvalue spectrum of the QCD Dirac operator and from the topological susceptibility 
using lattice techniques. I conclude the discussion with the possible consequences of presence of such non-trivial 
topological structures at these temperatures for the experiments.

\section{$U_A(1)$ puzzle and a way towards its resolution}
Effects of anomalous $U_A(1)$ symmetry on QCD phase transition  with two light quark flavors is  known from model 
quantum field theories with same symmetries as QCD. However the $U_A(1)$ breaking is implemented through an effective 
term in the Lagrangian whose strength can only be determined from first principles lattice QCD studies. However 
lattice studies have their own share of systematic uncertainties. One needs to choose a sufficiently large lattice 
volume to contain enough of these topological structures. Second and the most crucial aspect is to control the lattice cut-off effects. 
Most lattice fermion discretizations break the continuum chiral symmetry  and the flavor singlet $U_A(1)$ explicitly. Only two known lattice 
fermion discretizations, the overlap and the domain wall fermions in the limit of infinite large fifth dimensional extent 
on which it is defined, are known to realize chiral symmetry explicitly on the lattice. These operators also satisfy an exact 
index theorem, thus the $U_A(1)$ breaking can be related to the non-trivial gauge field topology. 
\begin{figure}[h]
\begin{center}
\includegraphics[scale=0.4]{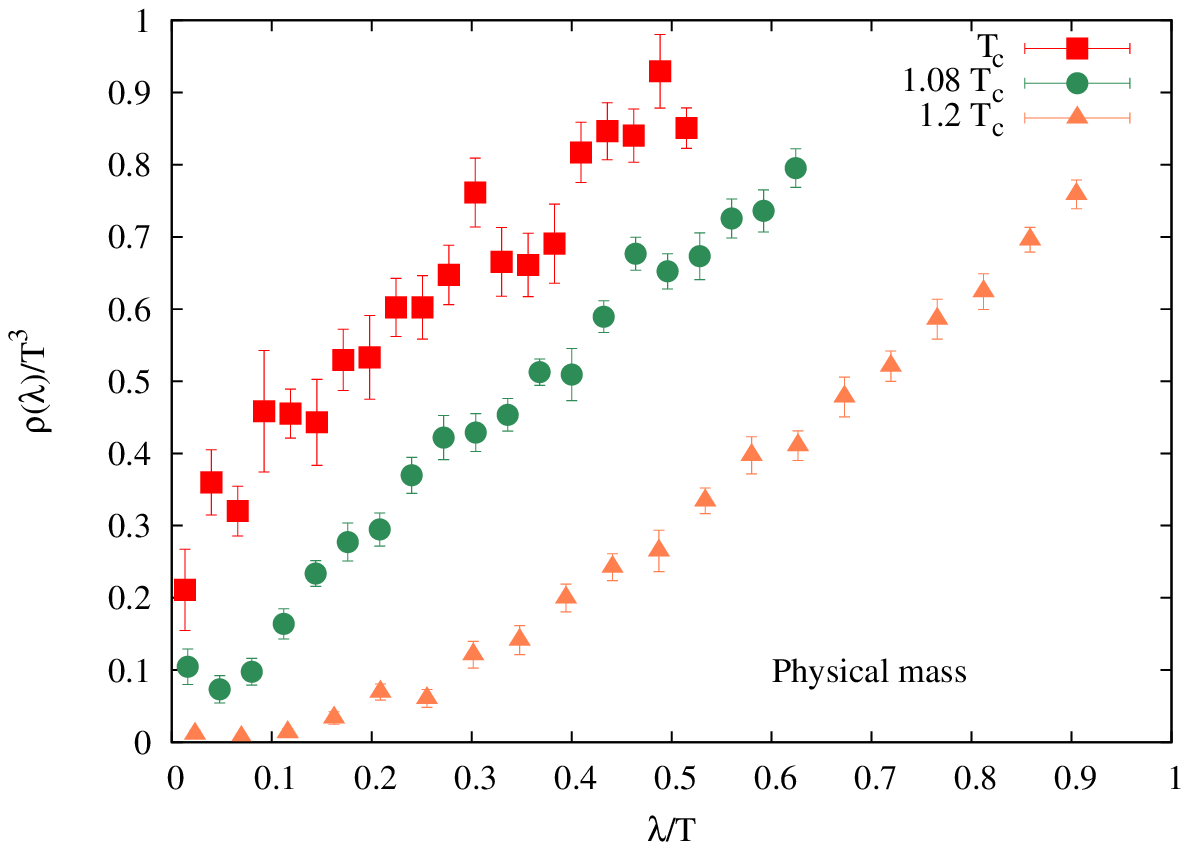}
\includegraphics[scale=0.4]{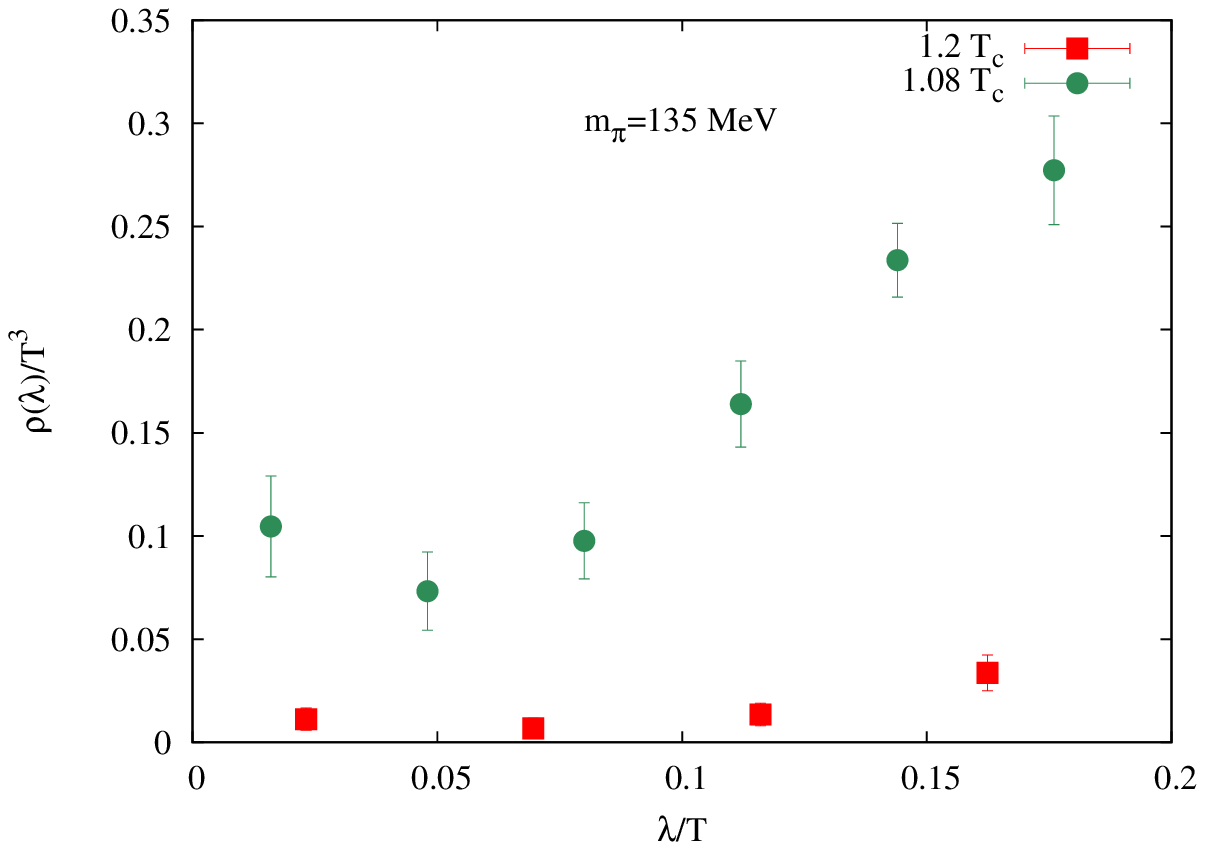}
\caption{The eigenvalue spectrum of QCD Dirac operator using overlap fermions (left panel) on domain wall gauge 
configurations. The near-zero part the spectrum (right panel) is shown as a function of temperature for physical quark mass. }
\label{fig:dweig}
\end{center}
\end{figure}
We use the gauge configurations generated with  M\"{o}bius domain wall fermion discretization and Iwasaki gauge action with 
dislocation suppressing determinant. The details of the gauge configurations are in ~\cite{dw}. The light and the strange quark masses are 
physical such that $m_{u,d}<<\Lambda_{QCD}$. The small residual chiral symmetry breaking for these domain wall fermion 
configurations is $\mathcal{O}(10^{-4})$ and the index is not exactly realized. Hence we use overlap fermions as valence 
quarks to count the zero modes of the gauge configurations and measure the eigenvalue spectrum. The overlap operator 
$D_{ov}=M\left[1+\gamma_5 ~\mathrm{sgn}\left(\gamma_5 ~D_W(-M)\right)\right]$, consists of a sign function of Wilson Dirac operator $D_W$, 
which is the lattice discretization of the continuum  Dirac operator with an irrelevant term and a domain wall height parameter 
$M$. The sign function was measured to a precision of about $10^{-10}$ and the violation of the exact chiral symmetry given by 
the Ginsparg-Wilson relation was of the same order of magnitude for each gauge configuration.  On each configuration, $50$ eigenvalues 
of $D_{ov}^\dagger D_{ov}$ were measured using the Kalkreuter-Simma Ritz algorithm~\cite{ks}. We analyzed about $100-150$ configurations 
at each temperature. Our lattice volumes are sufficiently large enough, $V^{1/3}\geq 4/m_\pi$ essential for such a study. 
In the left panel of Fig.~\ref{fig:dweig} we show our results for the eigenvalue density of the QCD Dirac operator as a function 
of temperature. As the temperature increases one observes a gradual appearance of a small near-zero mode peak separating from 
the bulk eigenvalues whose contribution diminishes at $1.2~T_c$ as shown in the right panel of Fig.~\ref{fig:dweig}. One needs 
a careful study of the near-zero modes since its origin may be due to residual chiral symmetry breaking effects of the 
domain wall fermions~\cite{cossu}. 
 
\begin{figure}[h]
\begin{center}
\includegraphics[scale=0.4]{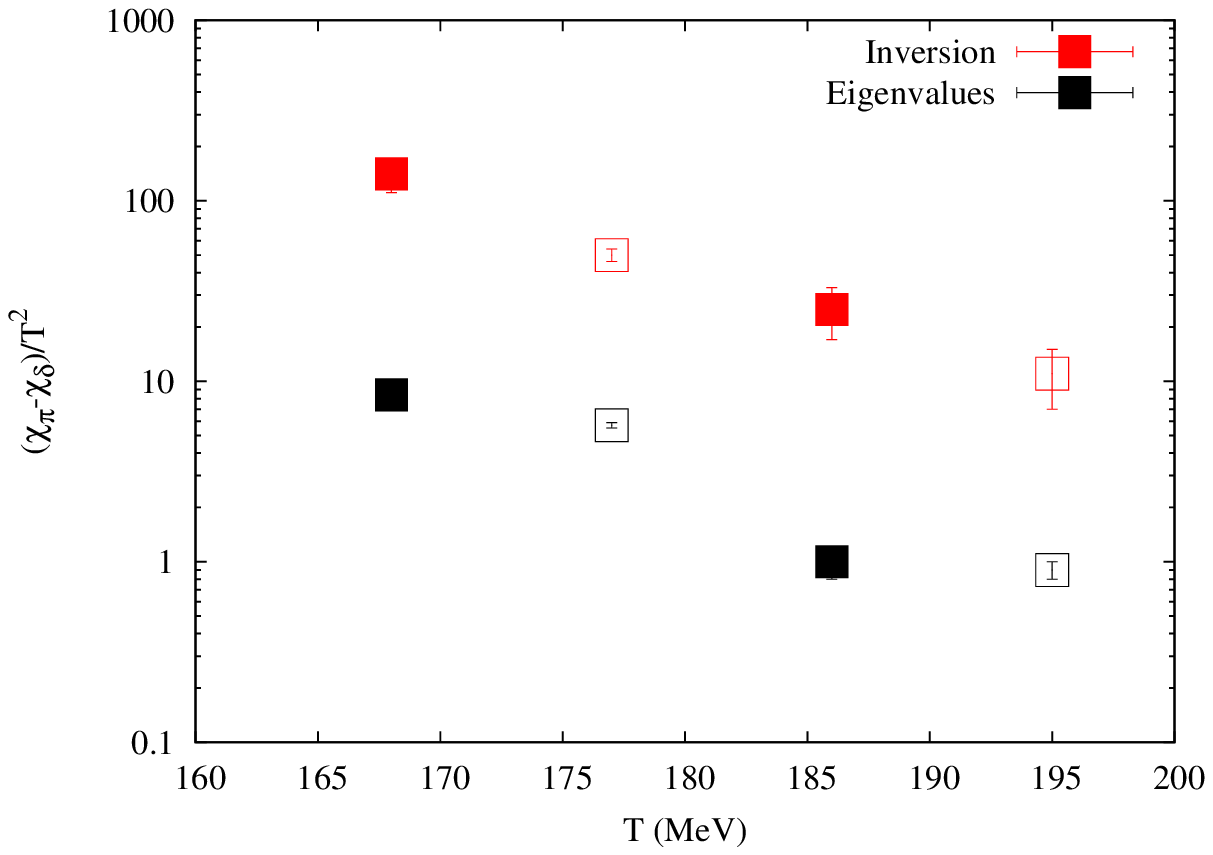}
\includegraphics[scale=0.4]{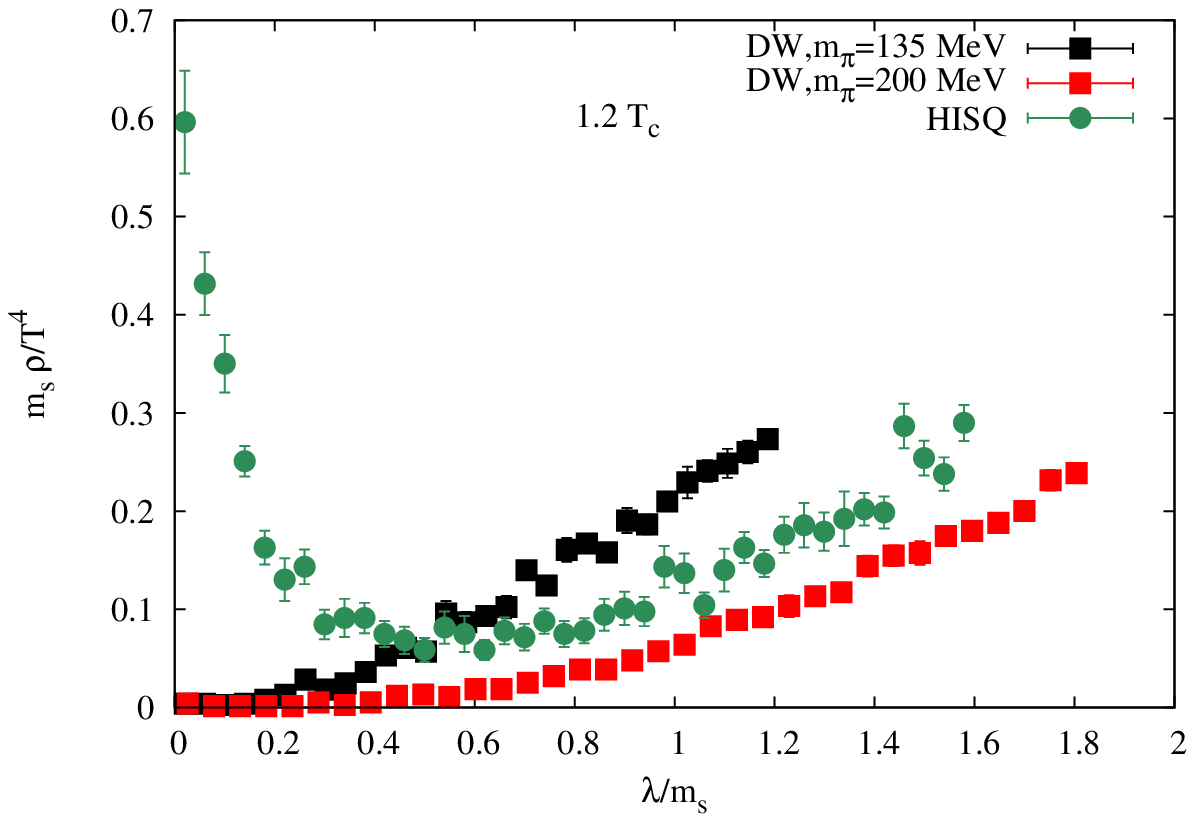}
\caption{The $U_A(1)$ breaking in the scalar-pseudoscalar sector measured through the observable 
$(\chi_\pi-\chi_\delta)/T^2$ (left panel) shows that $U_A(1)$ is not effectively restored at $1.2~T_c$. 
The black data points are the contribution from the lowest 50 eigenvalues to the total value 
denoted by the red points taken from Ref.~\cite{dw}. In the right panel, the renormalized eigenvalue 
spectrum is plotted for different fermion discretizations show quite a robust behavior for the bulk part.}
\label{fig:chipd}
\end{center}
\end{figure}

One of the measures of $U_A(1)$ breaking is the non-vanishing value of the difference between integrated pion and delta meson 
correlation functions, $(\chi_\pi-\chi_\delta)/T^2$. From the left panel of Fig.~\ref{fig:chipd} it is evident that this quantity is non-zero 
at even $1.2~T_c$ and a large contribution comes from the first $50$ eigenvalues. It is also known that if the analytic(bulk) 
part of the eigenvalue spectrum goes as $\rho(\lambda)\sim\lambda^3$, then $U_A(1)$ breaking effects are invisible in 
upto 6-point correlation functions in the scalar and pseudo-scalar sectors~\cite{aoki}. We observe that the analytic part of the 
renormalized eigenvalue spectrum has a behavior quite different from $\lambda^3$ at $1.2~T_c$ irrespective of the lattice 
discretization( right panel of Fig.~\ref{fig:chipd}), by comparing our eigenspectrum with those previously measured by us using 
improved staggered (HISQ) fermions~\cite{viktor}. We conclude that  $U_A(1)$ is not effectively restored at 
$1.2~T_c$ and the contribution of the analytic part of the eigenspectrum to its breaking is quite robust unaffected by 
lattice cut-off effects. For more details see \cite{viktordw}.

\section{The topological structures in strongly coupled QCD at finite temperature}
\begin{figure}[h]
\begin{center}
\includegraphics[scale=0.4]{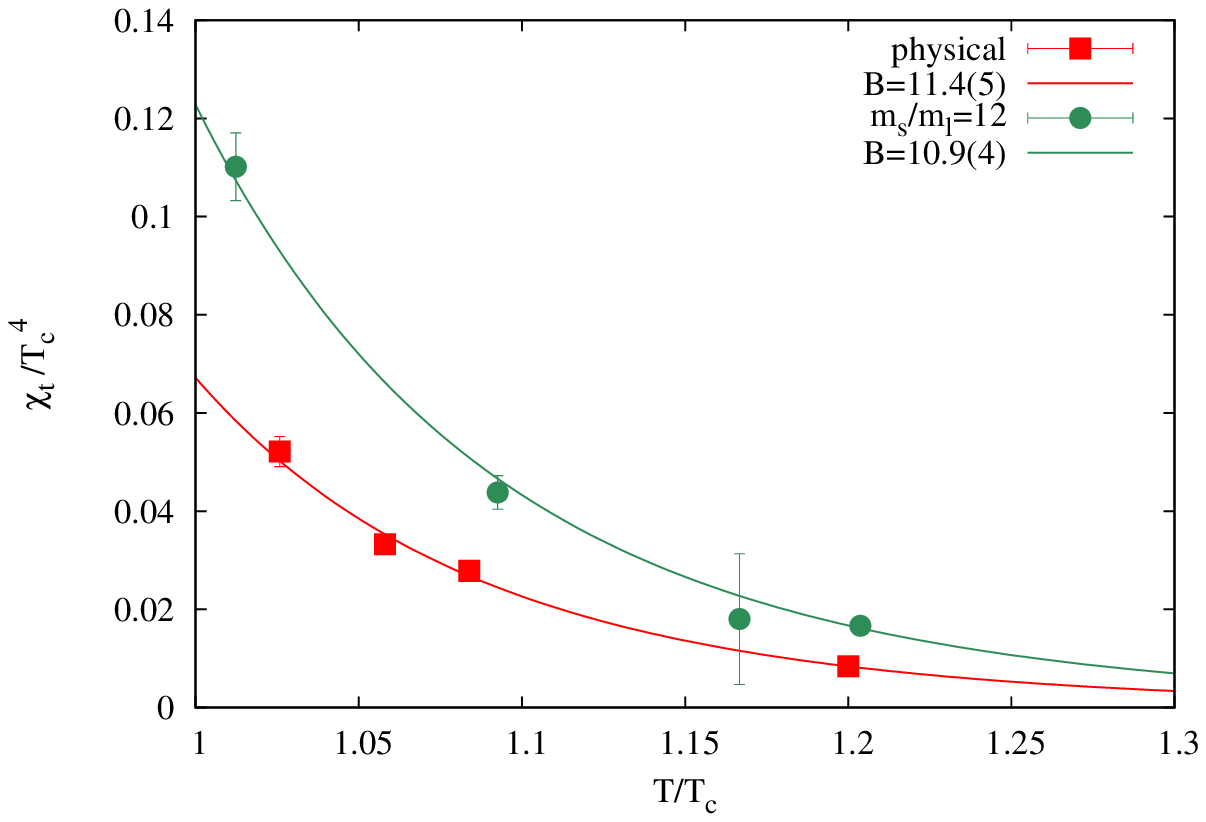}
\includegraphics[scale=0.4]{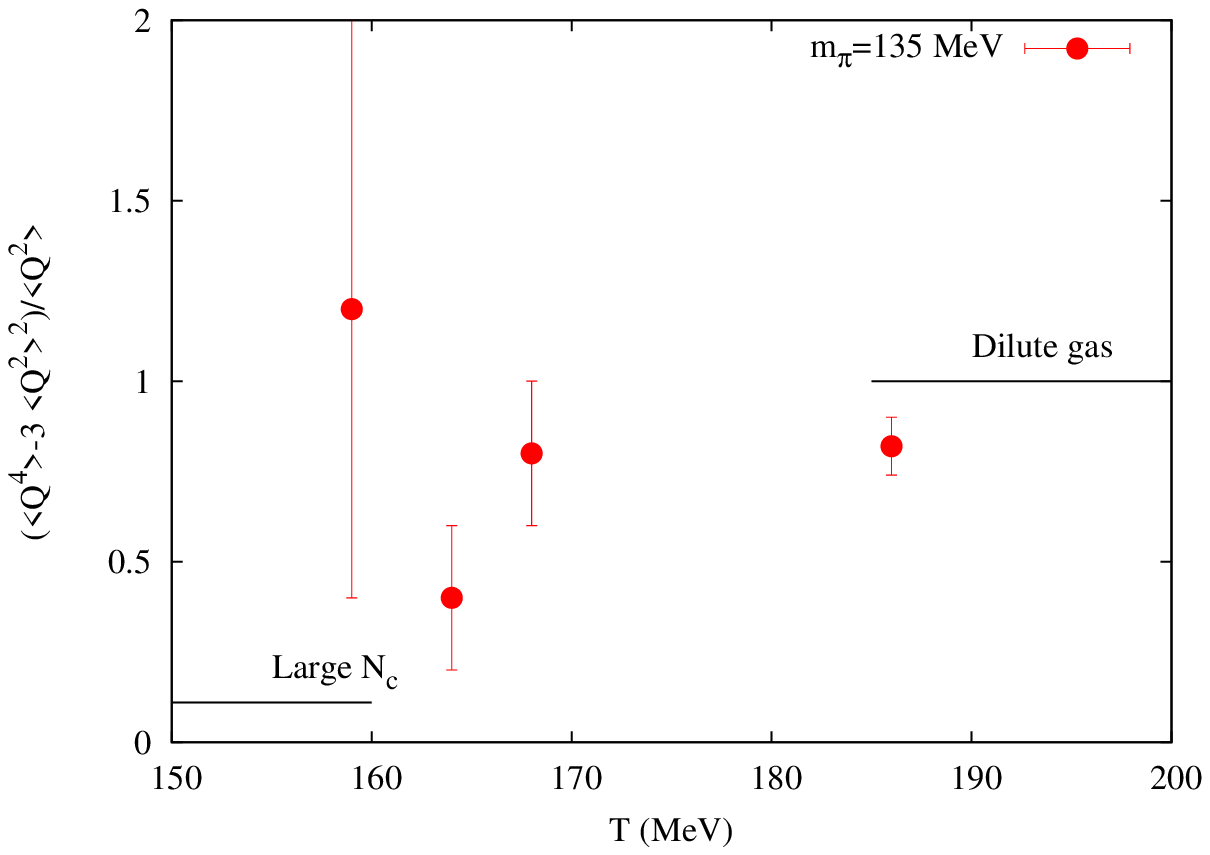}
\caption{The temperature dependence of topological susceptibility (left panel) shown as a function of quark mass and the 
second moment of the topological susceptibility for physical quark mass (right panel).}
\label{fig:topsusc}
\end{center}
\end{figure}

The $U_A(1)$ breaking at $T=0$ leads to a finite $\eta^{'}$ mass and is related to the physics of 
instantons~\cite{thooft}. Near the chiral crossover, the QCD vacuum properties are 
not yet well understood. Instanton liquid model~\cite{shuryak} provides explanation for chiral symmetry 
breaking and properties of hadronic correlations~\cite{shuryakilg1}. It is also known that QCD instantons at finite 
temperature with non-trivial holonomy has substructures called dyons~\cite{kvbaal} which carry both 
electric and magnetic charges. In order to understand the existence and interactions between these different 
topological structures, we measure the topological susceptibility in QCD, defined as 
$\chi_t= \frac{T}{V}\left[<Q^2>- <Q>^2\right]$. It measures the local fluctuations
of the topological charge $Q$ though $\langle Q\rangle=0$ in the thermodynamic limit. Since we use overlap fermions with 
an exact index theorem, we for the first time, report the results for $\chi_t$ in lattice QCD with exact chiral symmetry at high temperatures. 
The $\chi_t$ shows sensitivity to the quark mass as seen in the left panel of Fig.~\ref{fig:topsusc}. For light quark masses 
about twice the physical values,  $\chi_t$ is enhanced by factor two near $T_c$.  Fitting  $\chi_t/T_c$ to a function 
$A\times T^{-B}$, we get an exponent $B=11.4(5)$ for physical quark masses and $B=10.9(4)$ for $m_s/m_l\sim 12$. For a dilute 
gas of instantons,  including  first order quantum fluctuations over the classical gauge configurations gives 
temperature dependence of $\chi_t\sim T^{-7}$.   From the temperature dependence of our data for $\chi_t$, it is evident 
that dilute gas picture has not yet set in at $1.2~T_c$. It also gives a hints about possible existence of dyons 
in the QCD medium just above $T_c$, though the trend could be also consistent with the instanton liquid model. 
A still better probe is to look into the higher moments of topological charge fluctuations. In the right panel of 
Fig.~\ref{fig:topsusc}, we show our results for  $\frac{<Q^4>-3 <Q^2>^2}{<Q^2>}$. While near $T_c$, the data is consistent 
with an interacting instanton ensemble that scales as $N_c$, the dilute instanton gas limit do not set in even at 
$1.2~T_c$. There are strong residual interactions between the finite temperature instantons or they dissociate into 
dyonic components. This is consistent with our earlier observation that dilute gas limit may set in only around 
$1.5~T_c$~\cite{viktor}.

\section{Implications for experiments and outlook}
Our study has comprehensively shown that $U_A(1)$ for QCD with two light quark flavors is explicitly broken at $1.2~T_c$. 
Though the near-zero part of the QCD Dirac eigenspectrum still needs to be studied in detail, the contribution of the analytic 
(bulk) part of the spectrum to $U_A(1)$ 
breaking is quite robust insensitive to lattice cut-off effects. We also looked into the topological constituents that contribute 
dominantly to $\chi_t$ just above $T_c$. We speculate about the presence of dyons from our data. It has been recently reported that 
magnetic monopoles may explain the jet-quenching phenomena in the 
low $p_T$  region in the range  $3~$GeV$<p_T<20$ GeV~\cite{miklos}.  Dyons with non-zero magnetic charges could also possibly 
contribute to the jet quenching phenomenology in the strongly coupled QGP. Our first results for $\chi_t$ in 
the region near $T_c$, also opens up the possibility of accounting for the chiral imbalance in anomalous hydrodynamics through 
$\chi_t$ as an input instead of a non-zero axial chemical potential.

$\mathbf{Acknowledgements}$
The work has been partially supported  through the  DOE  under  Contract  No.   de-sc0012704 and 
the BMBF under grants  05P12PBCTA and 05P15PBCAA. The GPU codes used in our work were in part based on some publicly 
available QUDA libraries~\cite{quda}.












\end{document}